\documentclass[letterpaper, 10 pt, conference]{ieeeconf}  
\usepackage{graphicx}      
\usepackage{float}	
\usepackage{tikz}
\usepackage{amsmath}
\usepackage{amsfonts}
\usepackage{amssymb}
\usepackage[ruled, vlined]{algorithm2e}
\newtheorem{remark}{Remark}
\newtheorem{theorem}{Theorem}

\IEEEoverridecommandlockouts                              

\begin{document}

\title{\LARGE \bf
Loewner-based Data-driven Iterative Structured Control Design
}

\author{Basile Bouteau$^{1}$ and Pauline Kergus$^{2\star}$ and Pierre Vuillemin$^{1}$
\thanks{$^{1}$ ONERA, Toulouse, France }
\thanks{$^{2}$ Department of Automatic Control, Lund University, Sweden}%
\thanks{$^{\star}$ Corresponding author: \tt \small pauline.kergus@control.lth.se}
}

\maketitle
\thispagestyle{empty}
\pagestyle{empty}

\tikzstyle{sum}=[draw,circle,text width = 0.3cm]
\tikzstyle{block}=[rectangle, draw=black, text centered]

\newcommand*{\new}{\textcolor{black}}

\begin{abstract}                
Stability enforcement remains a challenge in data-driven control paradigms, \new{where no parametrised model of the system is available}. \new{In \cite{kergus2019filtrage} for instance, the system's instabilities are estimated in order to enforce a closed-loop stability constraint on the controller reduction step. In order to avoid this preliminary estimation of instabilities, this paper proposes to embed a closed-loop stability constraint in the design. To that extent, an optimization problem is formulated in order to improve matching between the reference model and the closed-loop while maintaining internal stability. The proposed iterative procedure to solve this problem is illustrated on two numerical examples.}
\end{abstract}



\section{Introduction}
For many practical applications, a model cannot be derived from physical laws and input/output data may be the only accessible information concerning the system. In these cases, for control purposes, system data can be used to identify a model of the system. Then, based on closed-loop specifications, a controller can be designed applying some model-based techniques. However, in some context, the model of the system can be too complex or too costly to obtain. It may then be easier and faster to try to derive a controller by combining directly the system data and the specifications, as highlighted in \cite{aastrom1995pid}. These methods are known as Data Driven Control (\textbf{DDC}), \new{see \cite{hou2013model} for an overview}. Even if the traces of these methods go back to the 40s with the PID-tuning-method of \cite{ziegler1942optimum}, \textbf{DDC} methods have \new{regained interest in the control community} in the past 25 years \new{due to the profusion of data that is now available}. 

\new{Among \textbf{DDC} techniques, an appealing approach is the model reference framework. The objective is to design a controller such that the closed-loop is as close as possible to a reference model specified by the user. For instance,} the Iterative Feedback Tuning (\textbf{IFT}, \cite{hjalmarsson2002iterative}) is an online method where an optimal structured controller is obtained through an iterative process consisting in minimizing the error between the output and the desired one. The Correlation-based Tuning (\textbf{CbT}, \cite{karimi2002convergence}), is a time-domain method which consists in minimizing the correlation between the reference signal and the error between the closed-loop output and the desired output, over some class of controller. The Virtual Reference Feedback Tuning (\textbf{VRFT}, \cite{campi2002virtual}) is a one-shot off-line method. \new{From one set of input/output time-domain data, the parameters of the controller are obtained through convex optimisation to match the ``ideal" controller, defined according to the available data and the reference model. The ideal controller is the one that would give exactly the reference model and is a central concept in the reference model strategy: in the end, the resulting controller should approach the ideal controller as much as possible.} 

\new{While these approaches sound very appealing to a user since they only require data and a reference model, they rely on the assumption that the ideal controller can be reached by the chosen controller structure. Therefore the choice of the controller parametrization becomes a critical choice that can be hard to make without a plant's model. To that extent, the Loewner Data-Driven Control (\textbf{L-DDC}, \cite{kergus2017frequency}) proposes to identify a reduced-order controller from the ideal controller's frequency response samples that are computed from the reference model and frequency-domain data from the plant.}

\new{Another drawback of the reference model strategy, common to all the mentioned techniques, is that the reference model should be carefully chosen to be achievable by the system. Indeed, a reference model that does not respect the fundamental performances limitations of the system leads to an ideal controller that compensate the plant's instabilities \cite{bazanella2011data}. The ideal case is therefore internally unstable and such behaviour should not be pursued as an objective. Closed-loop stability and the choice of the reference model therefore appear to be strongly linked. To tackle these issues,} additional conditions on the controller have been taken into account for \textbf{IFT} in \cite{de1999iterative} and \textbf{CbT} in \cite{karimi2007non}. \new{For the \textbf{VRFT}, \cite{bazanella2011data} and \cite{selvi2018towards} have proposed to parametrize the reference model function and to find a good one along with the design of the controller. In \cite{van2009data}, it is proposed to define the reference model according to the nature of the system and through a stable filter that will determine the performances. The filter is applied to functions that are already known to be achievable, thanks to an initial stabilizing controller for unstable plants for instance.} Concerning the \textbf{L-DDC}, \new{the available data are first analyzed to detect and estimate instabilities before building an achievable reference model \cite{kergus2019filtrage}. A stability constraint based on the small gain is then applied to ensure closed-loop stability during the controller reduction, similarly to \cite{van2009data}.}

\new{However, regarding the \textbf{L-DDC}, the detection and estimation of instabilities require some expertise, mainly in the multivariable case.} The main contributions of this paper consist in proposing a structured controller synthesis counterpart of the \textbf{L-DDC}, leading both to internal stability and closed-loop performances similar to a reference model, without any need of previous unstable poles/ zeros identification. This synthesis is formulated as an iterative procedure, solving optimization sub-problems where the internal stability criterion is based on the application of the small-gain theorem. To solve these sub-problems, a one-shot data-based estimation of the $\infty$-norm is suggested through the use of Loewner Framework. \new{Compared to the traditional \textbf{L-DDC} framework, controller structuring allows to avoid unquantified loss of performances during controller reduction but also prevents from compensating resonant modes of the system, which is known to be non robust.}

This paper is organized as follows: in Section \ref{Problem formulation}, the data-driven structured controller design is formulated as an optimization problem and the key tools involved in its resolution are presented. The iterative procedure proposed in this paper is then introduced in Section \ref{Contributions}. In Section \ref{Simulation results}, the proposed algorithm is applied on two numerical examples. The first one is a DC motor, for which the ideal controller is reachable, and serves as a proof of concept. Then, the algorithm is applied to a multivariable and non-minimum phase aeronautic example to check the performances reached in the mismatch case, when the reference model  is unreachable while ensuring internal stability and/or the chosen controller structure). Finally, conclusions and outlooks are given in Section \ref{Conclusion}.

\section{Problem formulation and key elements}
\label{Problem formulation}

\begin{figure}
    \centering
    \includegraphics[width=0.4\textwidth]{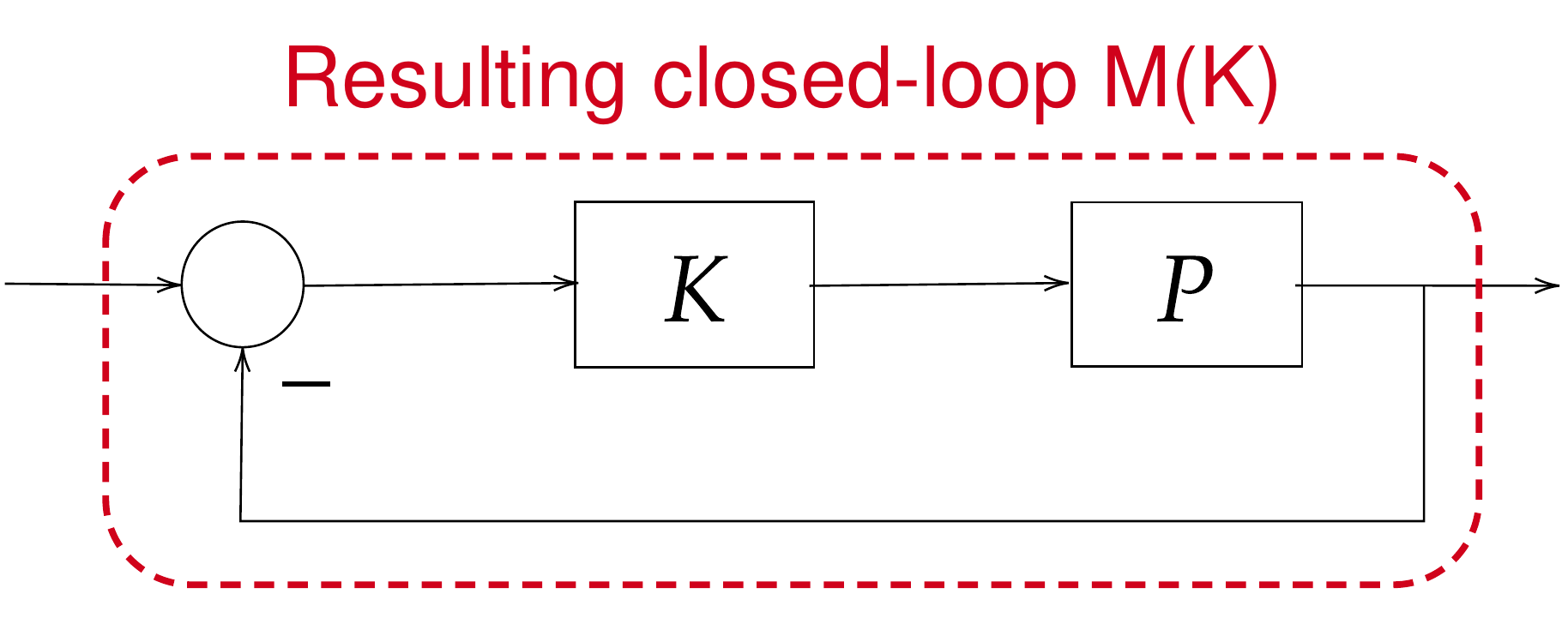}
    \caption{Considered control interconnection.}
    \label{fig:control_interconnection}
\end{figure}

A classic feedback interconnection is considered in this paper, as visible on Figure \ref{fig:control_interconnection}. Well known performance limitations are imposed by the unstable poles and NMP zeros of the plant \cite{havre1997limitations} thus making some reference models unachievable in terms of internal stability. Therefore, a challenge is to find a controller that leads both to internal stability and to closed-loop performances close to the reference model. As in \textbf{L-DDC}, the proposed technique is based on one set of frequency-domain data $\left(j\omega_k,\Phi_k\right)_{k=1...N}$. 

The controller design is formulated here as the following optimization problem
\begin{equation}
  \mathcal{P}:\left \lbrace
   \begin{array}{rc}
     \displaystyle \min_{\mathbf{K} \in \mathcal{K}} & \Vert\mathbf{M}_d-\mathbf{M}(\mathbf{K})\Vert \\
     s.t.&
     \begin{array}[t]{l}
          \text{Closed-loop system internally stable}\\
         \mathbf{K} \text{ Stable}
         \end{array}
   \end{array}
\right.,
\label{P}
\end{equation}
where $\mathbf{M}_d$ is the reference model specified by the user representing the desired closed-loop behaviour and $\mathbf{M}(\mathbf{K})$ is the closed-loop transfer obtained with the controller $\mathbf{K}$, $\Vert.\Vert$ is some norm to be specified and $\mathcal{K}$ is a subspace of the controller space which only considers controllers with a chosen structure. \new{In many applications, it is preferred not to introduce any instabilities in the open-loop, therefore the stability of the designed controller is enforced.}

The objective function of problem $\mathcal{P}$ \eqref{P}, the internal stability constraint and the controller parametrization are detailed hereafter before presenting the proposed algorithm in Section \ref{Contributions}.

\subsection{Objective function: matching the reference model}
\label{obj}
In order to obtain the desired closed-loop performances, it is proposed in \eqref{P} to minimize the distance between the reference model $\mathbf{M}_d$ and the closed-loop transfer with a controller $\mathbf{K}$. This is done as follows:
\begin{equation}
    \Vert\mathbf{M}_d-\mathbf{M}(\mathbf{K})\Vert=\frac{1}{N}\sum_{k=1}^N\Vert\mathbf{M}_d(\imath\omega_k)-\mathbf{M}(\mathbf{K},j\omega_k)\Vert_{\mathcal{F}}^2,
    \label{norm}
\end{equation}
where $\Vert.\Vert_{\mathcal{F}}$ is the Frobenius norm. The closed-loop transfer $\mathbf{M}(\mathbf{K})$ can be computed at each sample point as follows:
\begin{equation}
    \mathbf{M}(\mathbf{K},\imath\omega_k)=\left(I+\Phi_k\mathbf{K}(\imath\omega_k)\right)^{-1}\Phi_k\mathbf{K}(\imath\omega_k).
    \label{MKsamples}
\end{equation}

This norm quantifies the quadratic error between the reference model and the closed-loop at the frequency sample where plant's data is available. Should the points be uniformly distributed and their amount tending to $\infty$, equation \eqref{norm} would tend to $\Vert\mathbf{M}_d-\mathbf{M}(\mathbf{K})\Vert_{2}^{2}$.

\subsection{Internal stability constraint}
\begin{figure}
    \centering
    \includegraphics[width=0.4\textwidth]{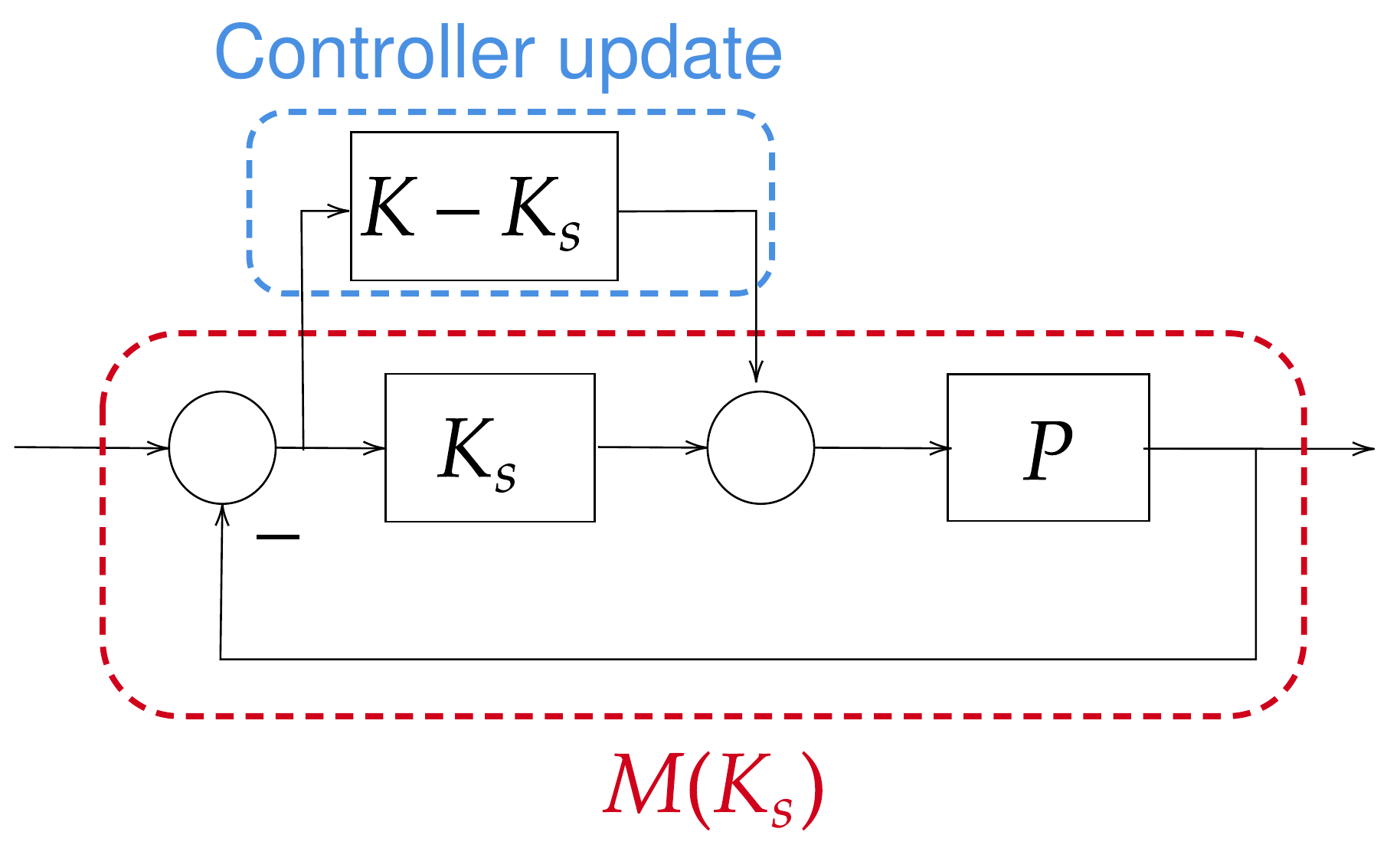}
    \caption{Equivalent interconnection to the resulting closed-loop visible on Figure \ref{fig:control_interconnection} based on a stabilizing controller $\mathbf{K}_s$, allowing to use the small-gain theorem.}
    \label{fig:small_gain}
\end{figure}

Enforcing closed-loop internal stability is an important feature in the proposed approach. As the plant is known only through some input/output data, closed-loop internal stability must be assessed in a data-driven way. To that extent, we assume that a stabilizing controller $K_s\in \mathcal{RH}_\infty$ is available. This assumption will be discussed in Section \ref{Contributions} and will be valid along the iterative procedure that is proposed in this paper. As in \cite{van2009data} and \cite{kergus2019filtrage}, a sufficient stability condition is derived from the small-gain theorem. It relies on rewritting the closed-loop visible on Figure \ref{fig:control_interconnection} as the one visible on Figure \ref{fig:small_gain}. The internal stability constraint is given in Theorem \ref{small_gain_thm_stab}.

\begin{theorem}
Let $\mathbf{G}=\mathbf{P}(1-\mathbf{M}(\mathbf{K}_s))$ and $\gamma>0$. Then the interconnected system shown on Figure \ref{fig:small_gain} is well-posed and internally stable for all stable $\mathbf{\Delta}=\mathbf{K}-\mathbf{K}_s$ with:
    \begin{itemize}
        \item[(a)] $\left\Vert \mathbf{\Delta} \right\Vert_\infty \leq \frac{1}{\gamma}$ if and only if $\left\Vert \mathbf{G} \right\Vert_\infty <\gamma$
        \item[(b)] $\left\Vert \mathbf{\Delta} \right\Vert_\infty < \frac{1}{\gamma}$ if and only if $\left\Vert \mathbf{G} \right\Vert_\infty \leq\gamma$
    \end{itemize}
    \label{small_gain_thm_stab}
\end{theorem}

As the model of the plant $\mathbf{P}$ is unknown, the determination of the $\infty$-norm of $\mathbf{G}$ cannot be performed in a classic way. Data-based estimation of the $\infty$-norm are available in \cite{rojas2012analyzing} or via expert advice in \cite{rallo2017data}. Another approach based on Loewner interpolation is proposed hereafter.

\subsection{Loewner-based $\mathcal{H}_\infty$-norm estimation}
\label{Loewner}
The Loewner framework \cite{mayo2007framework} is traditionnaly known for model order reduction. While the \textbf{L-DDC} algorithm uses it for controller reduction, the proposed approach uses its ability to identify quickly a minimal realization of a transfer function $\mathbf{G}$ in order to estimate $\Vert \mathbf{G}\Vert_\infty$. This will be of importance to ensure closed-loop internal stability in a data-driven way, as highlighted in the previous paragraph.

\new{In the present case, the frequency response of $\mathbf{G}$ is available at frequencies where the plant response is known. Therefore, one wants to obtain a rational model $\hat{\textbf{G}}$ that satisfies the following interpolatory conditions:}

\begin{equation}
    \forall k=1\dots N, \hat{\mathbf{G}}(j\omega_k)=\mathbf{G}(j\omega_k).
    \label{CI}
\end{equation}

In order to construct such a representation, let us divide the sample points into two sets of interpolation points: 
\begin{equation*}
    \left\{j\omega_k\right\}_{k=1}^N=\left\{\mu_i\right\}_{i=1}^m\bigcup \left\{\lambda_j\right\}_{j=1}^p
\end{equation*}
From these, Loewner and shifted-Loewner matrices can be built as
\begin{equation}
    \begin{array}{c}
         \left[\mathbb{L}\right]_{i,j}=\frac{\mathbf{G}(\mu_i)-\mathbf{G}(\lambda_j)}{\mu_i-\lambda_j}\\
         
         \left[\mathbb{L}_\sigma\right]_{i,j}=\frac{\mu_i\mathbf{G}(\mu_i)-\lambda_j\mathbf{G}(\lambda_j)}{\mu_i-\lambda_j}  
    \end{array}.
\end{equation}
Then, a Singular Value Decomposition (SVD) is performed on the Loewner pencil $\left(\mathbb{L},\mathbb{L}_{\sigma}\right)$ such that,
\begin{equation}
    \begin{array}{lcr}
         \left[\mathbb{L},\mathbb{L}_{\sigma}\right]=Y_1\Sigma_1X_1^H&,&\left[\begin{array}{c}
                \mathbb{L}\\
               \mathbb{L}_\sigma
         \end{array}\right]=Y_2 \Sigma_2 X_2^H  \\
    \end{array},
\end{equation}
where $^H$ represents the conjugate transpose. During this process, the McMillan order $r$ of the interpolation model is computed such that $\Sigma_1 \in \mathbb{R}^{r\times r}$ and $\Sigma_2 \in \mathbb{R}^{r\times r}$, more details are available in \cite{mayo2007framework}. 

Finally, a realization of the interpolation model is given in a descriptor form by:
\begin{equation}
    \hat{\mathbf{G}}:\left\{ \begin{array}{r@{=}l}
        E\dot{x} & Ax+Bu \\
        y & Cx
    \end{array}\right.
\end{equation}
where the matrices are defined as follows:
\small
\begin{equation}
         E=-Y_1^H\mathbb{L}X_2,\quad A=-Y_1^H\mathbb{L}_\sigma X_2,\quad B=Y_1^H V,\quad C=WX_2,
\end{equation}\normalsize
where $V_i^T=\mathbf{G}(\mu_i)$ and $W_j=\mathbf{G}(\lambda_j)$. 
\begin{remark}
As proven in Chapter $4$ of \cite{antoulas2005approximation}, if $N$ noise-free samples are extracted from a rational model $G$ of order $n\leq N$, then $\hat{\mathbf{G}}$ is a realization of the entire system.
\label{loewner_realization}
\end{remark}

The obtained realization $\hat{\mathbf{G}}$ is used in the proposed approach to get an estimation of the $\mathcal{H}_\infty$-norm of the original system $\mathbf{G}$:
\begin{equation}
    \Vert \hat{\mathbf{G}} \Vert_\infty \approx  \Vert \mathbf{G} \Vert_\infty.
    \label{Hinf_approx}
\end{equation}
As highlighted by Remark \ref{loewner_realization}, in the noise-free case, a sufficient number of samples makes \eqref{Hinf_approx} a perfect approximation. Otherwise, if the data set is noisy or does not contain enough samples, $ \Vert \hat{\mathbf{G}} \Vert_\infty$ might underestimate $\Vert \mathbf{G} \Vert_\infty$.

\subsection{Controller parametrization}
\label{parametrization}

The missing part in order to fully describe the problem given in \eqref{P} is to describe the proposed controller structure. The objective is to have a parametrization as general as possible but also to enforce the stability of the controller easily.

To that extent, a ZPK-inspired structure allows to monitor easily the poles and, in the SISO case, the zeros of the controller. In addition, this choice of parametrization allows to fix the controller order. The structure is
\begin{equation}
    \mathbf{K}(s)=\frac{1}{d(s)}N(s),
\end{equation}
where $d(s)\in\mathbb{C}$ is a polynomial in $s$ and $N(s)\in \mathbb{C}^{n_i\times n_o}$ is a matrix of polynomials in $s$ ($n_i$ is the number of controller inputs while $n_o$ is the number of controller outputs, i.e. the number of command signals). The polynomial $d$ is parameterized by the real vector $\beta \in \mathbb{R}^{n_p}$ such that
\begin{equation}
    d(\beta,s)=\left(\prod_{l=1}^{\lfloor\frac{n_p}{2}\rfloor}\left(s^2+\beta_{2l-1}s+\beta_{2l}\right)\right)f(s+\beta_{n_p}),
\end{equation}
where,
\begin{equation}
    f(s+\beta_{n_p})=\left\{\begin{array}{cc}
         s+\beta_{n_p}& \text{if }n_p\text{ is odd}\\
         1& \text{if }n_p\text{ is even}
    \end{array}\right.,
    \label{def f}
\end{equation}
not to lose any generality.
In a similar way, each coefficient of the polynomial matrix $N_{i,j}(s)$ is structured with a vector $\alpha^{i,j}$ and the scalar $k^{i,j}$ such that, for all $1\leq i\leq n_i$ and $1\leq j\leq n_o$
{\small \begin{equation}
    N_{i,j}(\alpha^{i,j},s)=k^{i,j}\left(\prod_{l=1}^{\lfloor\frac{n_z^{i,j}}{2}\rfloor}\left(s^2+\alpha^{i,j}_{2l-1}s+\alpha^{i,j}_{2l}\right)\right)f(s+\alpha^{i,j}_{n_z^{i,j}}),
\end{equation}}
where $f$ is defined as in \eqref{def f}. The controller denoted $\mathbf{K}(\theta)$ in the sequence is thus parameterized by
\begin{equation}
    \theta=\left[\beta,\alpha^{1,1}...\alpha^{n_i,n_o},k^{1,1},k^{1,2}...k^{n_i,n_o}\right].
\end{equation}
One may notice that such a structure is general since all real controller can be reached with it. To ensure $\mathbf{K}(\theta) \in \mathcal{RH}_{\infty}$ the following constraint must be satisfied,
\begin{equation}
    \begin{array}{lc}
         \beta_l>0&  1\leq l\leq n_p \\
         n_p>n_z^{i,j}&  1\leq i \leq n_i,  1\leq j \leq n_o 
    \end{array}.
    \label{constraint_RHinf}
\end{equation}

This can be expressed in a compact form:
\begin{equation}
    A\theta<0
\end{equation}
where the matrix $A$ translating the constraints of \eqref{constraint_RHinf} is
\begin{equation}
    A=\left[\begin{array}{cc}
        -I_{n_p} & 0_{n_p \times \left(m-n_p\right)} 
   \end{array}\right].
\end{equation}

\section{Contribution: }
\label{Contributions}

Now that the key elements in the proposed problem $\mathcal{P}$ have been detailed, the proposed technique to solve it is introduced in this section. It relies on solving sub problems $\mathcal{P_i}$ in an iterative way. While the sub-problems $\mathcal{P}_i$ are built in the next paragraph \ref{subproblems}, the overall algorithm is presented in \ref{L-DISC}.

\subsection{Reformulation of the problem}
\label{subproblems}

Considering a stabilizing controller $\mathbf{K}_i$, the transfer $\mathbf{G}_i$ defined in Theorem \ref{small_gain_thm_stab} is known at the frequencies $\left\{\omega_k\right\}_{k=1}^N$ where plant's data $\left\{\Phi_k\right\}_{k=1}^N$ are available:
\begin{equation}
    \forall k=1\dots N, \mathbf{G}_i(j\omega_k)=\Phi_k(1-\mathbf{M}(\mathbf{K_i},j\omega_k)),
    \label{Gi_samples}
\end{equation}
where the samples $\mathbf{M}(\mathbf{K_i},j\omega_k)$ can computed as in \eqref{MKsamples}. As explained in \ref{Loewner}, interpolating this data set using the Loewner framework gives a minimal realisation $\hat{\mathbf{G}}_i$. An estimation of the $\mathcal{H}_\infty$ of $\mathbf{G}_i$ is then obtained:

\begin{equation}
    \gamma_i=\Vert \hat{\mathbf{G}}_i \Vert_\infty \approx \Vert \mathbf{G}_i \Vert_\infty.
\end{equation}

Following Theorem \ref{small_gain_thm_stab}, a controller $\mathbf{K}_{i+1}$ such that
\begin{equation}
    \begin{array}{c}
         \mathbf{K}_{i+1} \in \mathcal{RH}_{\infty}  \\
         \Vert\mathbf{K}_{i+1}-\mathbf{K}_i\Vert_\infty<\varepsilon\gamma_i^{-1} 
    \end{array},
\end{equation}
where $\varepsilon\leq1$, would also lead to internal stability.

\begin{remark}
The parameter $\varepsilon$ can be used as a safety for the user in case the data is noisy or may not be rich enough, meaning that there is a chance that $\gamma_i$ underestimates $\Vert \mathbf{G}_i \Vert_\infty$. However, decreasing $\varepsilon$ reduces the region around $\mathbf{K}_i$ to look for other stabilizing controllers, which may be problematic considering the conservatism of the small-gain theorem and of the subsequent sufficient internal stability condition.
\end{remark}

The controller structure, as defined in paragraph \ref{parametrization}, can then be used to define a proper optimization problem $\mathbf{P}_i$ as defined in \eqref{Pi}, according to a given stabilizing controller $\mathbf{K}_i$. The parameters $\theta_{i+1}$ solving $\mathcal{P}_i$ allows to find a stable controller $\mathbf{K}_{i+1}=\mathbf{K}(\theta_{i+1})$ that minimizes the objective function in a region centered on $\mathbf{K}_i$. Compared to $\mathbf{K}_i$, $\mathbf{K}_{i+1}$ results in a closed-loop $\mathbf{M}(\mathbf{K}_{i+1})$ closer to the desired reference model $\mathbf{M}_d$ given by the user, or at least at the same distance in the sense of the norm defined in paragraph \ref{obj}.

\begin{equation}
  \mathcal{P}_i:\left \lbrace
   \begin{array}{rc}
     \displaystyle \min_{\theta_{i+1} \in \mathbb{R}^{n_\theta}} & \frac{1}{N}\sum_{k=1}^N\Vert\mathbf{M}_d(\imath\omega_k)-\mathbf{M}(\mathbf{K(\theta_{i+1})},\imath\omega_k)\Vert_\mathcal{F}^2 \\
     s.t.&
     \begin{array}[t]{l}
         \Vert\mathbf{K}(\theta_{i+1})-\mathbf{K}_i\Vert_{\infty}<\varepsilon\gamma_i^{-1}\\
         A\theta_{i+1}<0
         \end{array}
   \end{array}
\right.
\label{Pi}
\end{equation}

\subsection{Loewner-based Data-driven Iterative Structured Control: the L-DISC algorithm }
\label{L-DISC}
The reformulation of the problem as $\mathcal{P}_i$, defined for a stabilizing controller $\mathbf{K}_i$, only allows to explore a given region around $\mathbf{K}_i$ to improve the closed-loop performances. The process is therefore repeated in order to explore a larger part of the chosen controller set. This is done by using $\mathbf{K}_{i+1}$, obtained when solving $\mathcal{P}_i$, as a new stabilizing controller to define the next problem $\mathcal{P}_{i+1}$.  

This procedure is called \textbf{L}oewner-based \textbf{D}ata-driven \textbf{I}terative \textbf{S}tructured \textbf{C}ontrol (\textbf{L-DISC}) and is summed up in Algorithm \ref{Algo}.

\begin{algorithm}
\SetAlgoLined
\SetKwInOut{Solution}{Solution}
\KwData{\begin{itemize}
    \item Samples of the frequency response of the system $\{\omega_{k},\Phi_k\}, \ i=k \ldots N$.
    \item Stable reference model $\mathbf{M}_d$.
    \item Initial stabilizing controller $\mathbf{K}_0$
    \item Controller structure $\mathcal{K}=\left\{\mathbf{K}(\theta), \theta \in \mathbb{R}^{n_\theta}\right\}$, see \ref{parametrization}
    \item Stop criteria $\eta$
\end{itemize}}
\Solution{
\begin{enumerate}
    \item Initialization:
    \begin{itemize}
        \item $i=0$
        \item $d_0=\Vert \mathbf{M}_d- \mathbf{M}(\mathbf{K}_{0})\Vert$
        \item $\Delta_0>\eta$
    \end{itemize}
    
    \item \textbf{While $\Delta_i>\eta$}
    \begin{enumerate}
        \item Evaluate $\left\{\imath\omega_k,\mathbf{G}_i(\imath\omega_k)\right\}$, see \eqref{Gi_samples}
        \item Use the Loewner framework to identify $\hat{\mathbf{G}}_i$ 
        \item $\gamma_i=\Vert\hat{\mathbf{G}_i}\Vert_{\infty}$
        \item Solve $\mathcal{P}_i$, see \eqref{Pi}, get the minimizer $\theta_{i+1}^\star$
        \item $\mathbf{K}_{i+1}=\mathbf{K}(\theta_{i+1}^\star)$
        \item $d_{i+1}=\Vert \mathbf{M}_d- \mathbf{M}(\mathbf{K}_{i+1})\Vert$
        \item $\Delta_{i+1}=d_i-d_{i+1}$
        \item $i\leftarrow i+1$
    \end{enumerate} 
\end{enumerate}
}
\caption{\textbf{L-DISC} algorithm}
\label{Algo}
\end{algorithm}

Regarding the convergence of the \textbf{L-DISC} algorithm, one may notice that $\theta_{i}^\star$ is feasible for the problem $\mathcal{P}_{i}$ since $A\theta_{i}^\star<0$ and $\Vert\mathbf{K}(\theta_{i}^\star)-\mathbf{K}_i\Vert_{\infty}=0<\frac{1}{\gamma_i}$. Consequently $d_{i+1} \leq d_{i}$. As a consequence, the sequence $(d_i)_{i\geq 0}$ is decreasing. Moreover, by definition, $d_i\geq 0,\forall i$. Therefore the sequence $(d_i)_{i\geq 0}$ is converging and one can show that
\begin{equation}
\begin{array}{ccc}
    \forall \eta>0 &\exists i_0\geq 0 &d_{i_0}-d_{i_{0}+1}\leq\eta .
\end{array},
\end{equation}
therefore ensuring that the \textbf{L-DISC} procedure terminates in a finite number of iterations.

\begin{remark}
\label{initialization}
An initial stabilizing controller $K(\theta_{init})$ can be obtained by solving the following problem initialized randomly:
\begin{equation}
    \mathcal{P}_{init}:\left \lbrace
   \begin{array}{rc}
     \displaystyle \min_{\theta_{init} \in \mathbb{R}^{n_\theta}} & \alpha(\hat{\mathbf{M}}(\mathbf{K}(\theta_{init}))) \\
     s.t.& A\theta_{init}<0
   \end{array}
\right.
    \label{stabilizing_K}
\end{equation}
where $\alpha(\cdot)$ denotes the spectral abscissa (maximum real part of the poles) of a transfer function and $\hat{\mathbf{M}}(\mathbf{K}(\theta_{init}))$ is a minimal realization of $\mathbf{M}(\mathbf{K}(\theta_{init}))$ obtained through the Loewner framework, similarly to what has been done in \ref{Loewner}. This approach will be used in the examples in the next section.
\end{remark}

\section{Numerical examples}
\label{Simulation results}
This Section illustrates the \textbf{L-DISC} procedure on two examples. The first one, a DC motor, should be viewed as a proof of concept: the model is given so that the true ideal controller can be compared with the obtained one. This example represents an ideal case since the controller structure allows to reach exactly the desired performances while preserving internal stability. On the other side, the second example illustrates the mismatch case: it consists in controlling the yaw and roll angle of a F16 air fighter, where the considered system is multivariable.

\subsection{An ideal case: the DC motor}
The considered system, a DC motor, is described by: 
\begin{equation}
    \frac{\Omega(s)}{U(s)}=\mathbf{P}(s)=\frac{\frac{K}{fR+K^2}}{\frac{JL}{fR+K^2}s^2+\frac{fL+JR}{fR+K^2}s+1}
    \label{DCmotor}
\end{equation}
where $\Omega$ is the angular velocity, $U$ is the input voltage, $K=0.021\,Nm.s/rad$ the electromagnetic coefficient, $f=0.0182\,Nm.s/rad$ the fluid friction coefficient, $R = 0.56\,Ohms$ the electrical resistance, $L = 5.63\,mH$ the inductance and $5\cdot10^{-4}\,kg^2.m^2$ the moment of inertia.

The model is stable and minimum phase. Therefore, any stable reference model respect the plant's fundamental limitations, see \cite{kergus2019filtrage}. The reference model is taken as:
\begin{equation}
        \mathbf{M}_d(s)=\frac{1}{\left(s/\omega_0\right)^2+2\xi s/\omega_0+1},
        \label{ref_model}
\end{equation}
where $\omega_0=10$ rad/s and $\xi=1$. Thus, considering \eqref{DCmotor} and \eqref{ref_model}, the controller leading exactly to the reference model is
\begin{equation}
    \mathbf{K}^\star(s)=12.618\frac{s^2+36.51s+4.011}{s(s+20)}.
\end{equation}
In order to be in an ideal case, the chosen controller structure is
\begin{equation}
    \mathbf{K}(\theta,s)=\theta_5\frac{s^2+\theta_3s+\theta_4}{s^2+\theta_1s+\theta_2},
\end{equation}
and it should be noted that $\mathbf{K}^\star=K(\theta^\star)$ with $$\theta^\star=[20 \ 0 \ 36.51 \ 4.011 \ 12.618].$$
In order to apply the proposed approach, $50$ samples logspaced between $10^{-2}$ rad/s and $10^2$ rad/s are extracted from the plant's model in \eqref{DCmotor}. The initial stabilizing controller $\mathbf{K}(\theta_{init})$ is obtained as explained in Remark \ref{initialization} and is defined by:
$$\theta_{init}=[0.2145 \ 0.1657 \ 0.5237 \ 0.2580 \ 0.8859].$$
The \textbf{L-DISC} procedure results in a controller $\mathbf{K}(\theta_f)$ where:
$$\theta_f=[15.7511 \ 0.1370 \ 25.5729 \ 2.9401 \ 0.14.3566],$$
resulting in the value $\Vert \mathbf{M}_d - \mathbf{M}(\mathbf{K}_f) \Vert = 8.5317 \cdot 10^{-6}$ for the objective function.

\begin{figure}
    \centering
    \includegraphics[width=\linewidth]{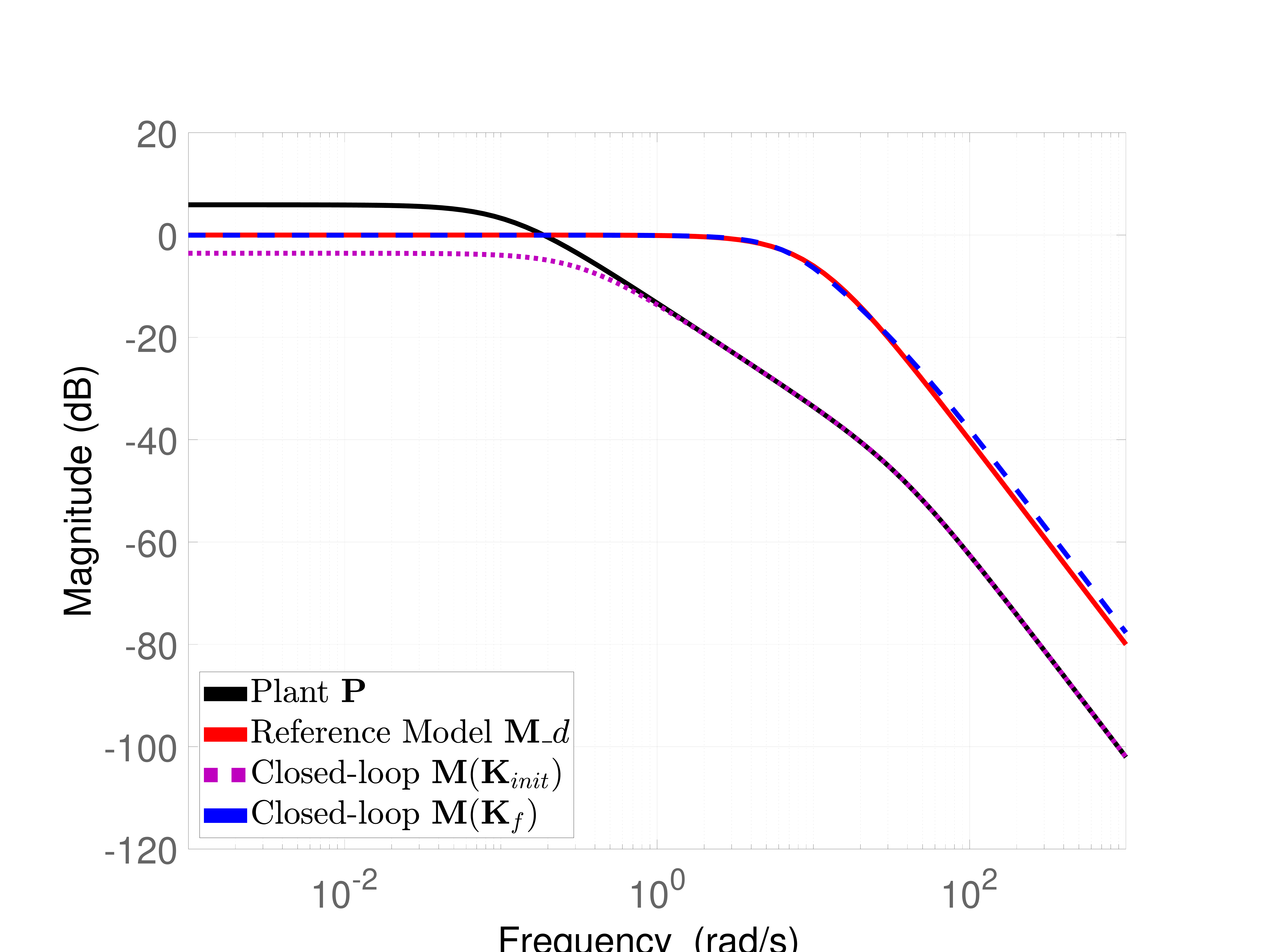}
    \caption{Closed-loop transfers in the DC motor case.}
    \label{fig:Frequency response of the controlled DC motor}
\end{figure}

The results are visible on Figure \ref{fig:Frequency response of the controlled DC motor}. In this ideal case, the frequency-response of the obtained closed-loop $\mathbf{M}(\mathbf{K}_f)$ matches exactly the one of the reference model $\mathbf{M}_d$, showing the ability of the \textbf{L-DISC} procedure to match a desired closed-loop behaviour.

\subsection{A mismatch case: roll and Yaw angle control on a F16 air-fighter}
\label{air fighter}
The yaw and roll angle of a F16 air-fighter is described in \cite{stevens2015aircraft} for an air speed of $62.5$ m/s and an angle of attack of $18.8$ deg by a NMP linearized model.  

\begin{figure}
    \centering
    \includegraphics[width=\linewidth]{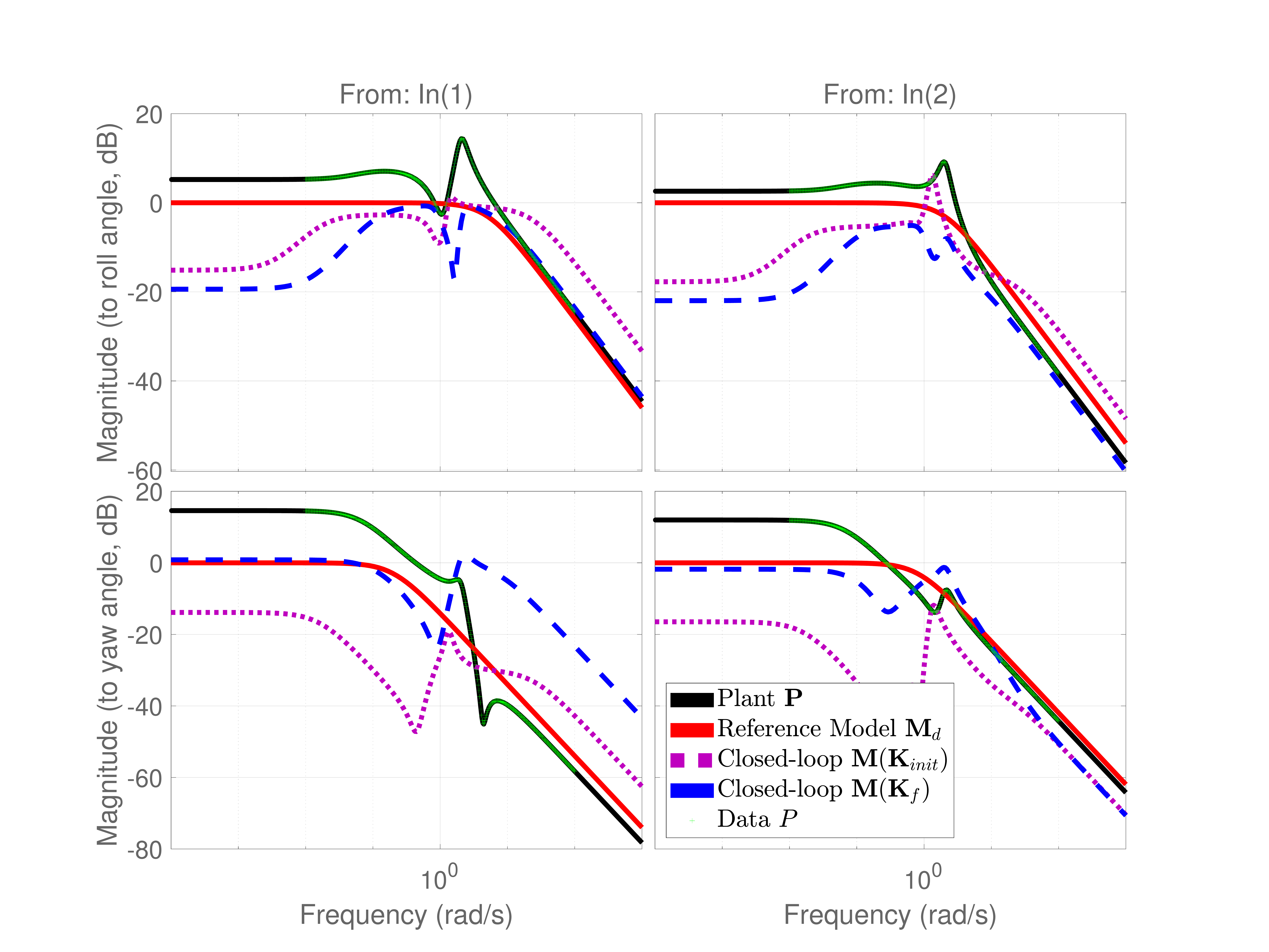}
    \caption{Closed-loop transfers for the angle control f16 air fighter}
    \label{fig:f16}
\end{figure}

In order to apply the proposed approach, $200$ samples logspaced between $10^{-2}$ rad/s and $10^2$ rad/s are considered, see Figure \ref{fig:f16}. The reference model is chosen such that the system is decoupled and such that the peaks on the diagonal transfers are limited without proceeding to high change of bandwidth (see Figure \ref{fig:f16}):
\begin{equation}
    \mathbf{M}_d(s)=\frac{1}{(s+5)(s+0.8)}\left[
    \begin{array}{cc}
         5(s+0.8)&0  \\
         0&0.8(s+5) 
    \end{array}\right].
\end{equation}
Note that $\mathbf{M}_d$ is minimum-phase while the plant's model is not, meaning that $\mathbf{M}_d$ cannot be reached under internal stability constraints: this is referred to as a mismatch case.

A MIMO controller structure of order $2$ where all transfer are bi-proper is considered. Considering the parametrization from \ref{parametrization}, the denominator $d(s)$ and each coefficient of the matrix $N(s)$ are of the form
\begin{equation}
     d(s)=s^2+\beta_1s+\beta_2,\quad
     N_{i,j}(s)=k_{i,j}\left(s^2+\alpha_1^{i,j}s+\alpha_2^{i,j}\right),
\end{equation}
thus leading to $14$ optimisation variables for $\theta$.
    
The initial stabilizing controller $\mathbf{K}(\theta_{init})$ is obtained as explained in Remark \ref{initialization}, see Figure \ref{fig:f16} for the frequency response of the associated closed-loop $\mathbf{M}(\mathbf{K}(\theta_{init}))$. Algorithm \ref{L-DISC} is then applied and results in a controller $\mathbf{K}(\theta_f)$ resulting in the value $\Vert \mathbf{M}_d - \mathbf{M}(\mathbf{K}_f)\Vert=$ for the objective function. 

\begin{figure}
    \centering
    \includegraphics[width=\linewidth]{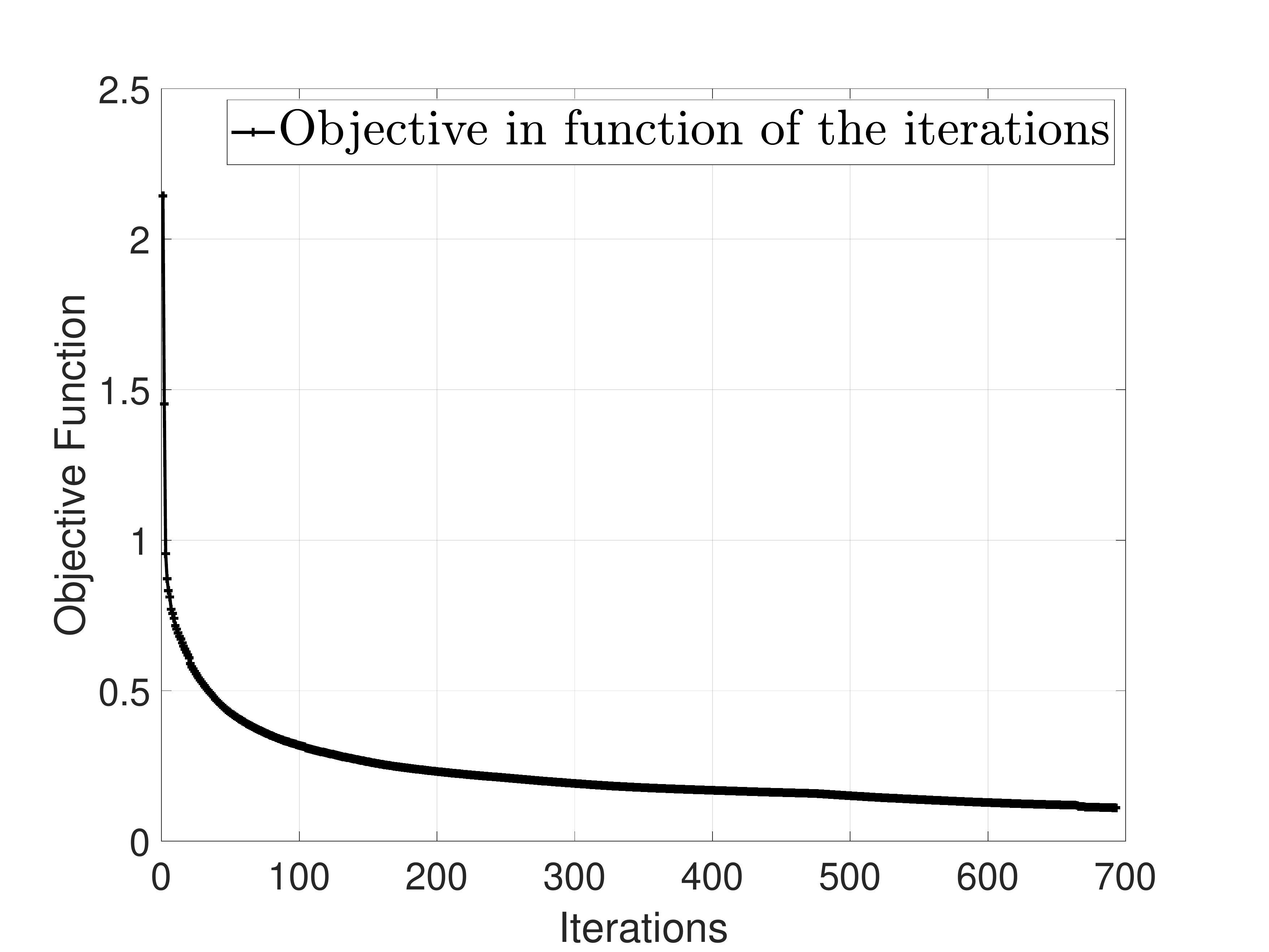}
    \caption{Evolution of the value of the objective functions according to the number of iterations.}
    \label{fig:evolution}
\end{figure}

The evolution of the objective function according to number of iterations is visible on Figure \ref{fig:evolution}: the objective function is only decreasing, as expected, and most significantly during the first iterations. However, after approximately $100$ iterations, the decrease is slower. A huge amount of iterations is needed to obtain a noticeable improvement of the objective function. This can be explained by the method used to solve the problem $\mathbf{P}_{i}$ \eqref{Pi}. Effectively, one may highlight that the optimization problem to solve is non-smooth, due to the $\infty$-norm constraint. Nevertheless, methods used for smooth optimization problem can still be used to reduce the objective function. It is actually not necessary to find a local/global optimum of the problem $\mathcal{P}_i$ to move to step $i+1$. But, when the frontier of the non-smooth feasible space are reached, it leads to a slow convergence. 

Finally, the closed-loop results are visible on Figure \ref{fig:f16}. In the considered frequency range, the closed-loop system has a similar behaviour than the reference model and the anti-diagonal transfer are maximized by $-20$ dBs which correspond to an attenuation of at least $90\%$. In the present case, the \textbf{L-DISC} algorithm allows to find a stabilizing controller of order 2 giving satisfying closed-loop performances compared to the desired reference model. It does not require to obtain a data-driven estimation the NMP zero of the model derived in \cite{stevens2015aircraft} while it would have been necessary through the \textbf{L-DDC} framework as detailed in \cite{kergus2019filtrage}.

\section{Conclusion}
\label{Conclusion}
In this paper, a data-driven technique based on frequency-domain data ensuring internal stability has been proposed. The specifications are given as a reference model. The proposed approach, denoted \textbf{L-DISC}, consists in optimizing the matching between the closed-loop and the reference model under a small-gain based constraint to ensure internal stability.

Compared to other approaches such as \cite{kergus2019filtrage}, the reference model only needs to represent the desired performances and does not need to be made achievable as in \cite{kergus2019filtrage}. This advantage is considerable since the modification of the reference model requires a deeper analysis of the plant's data to estimate its instabilities. Furthermore, the structured controller approach is a key advantage for practical applications where the controller form is not a tunable parameter. While, the iterative aspect of the procedure leads to higher computation time, it also makes the final closed-loop system less dependant on the initial stabilizing controller. 
A more efficient resolution of the sub-problems, based for instance, on convex relaxation and a possible weighting of the objective function with respect to the considered frequencies, is currently under investigation.


\bibliographystyle{unsrt}
\bibliography{biblio}

\begin{thebibliography}{10}

\bibitem{kergus2019filtrage}
P.~Kergus, M.~Olivi, C.~Poussot-Vassal, and F.~Demourant.
\newblock From reference model selection to controller validation: Application
  to loewner data-driven control.
\newblock {\em IEEE Control Systems Letters}, 2019.

\bibitem{aastrom1995pid}
K.~J. {\AA}str{\"o}m and T.~H{\"a}gglund.
\newblock {\em PID controllers: theory, design, and tuning}.
\newblock Instrument society of America, 1995.

\bibitem{hou2013model}
Z-S. Hou and Z.~Wang.
\newblock From model-based control to data-driven control: Survey,
  classification and perspective.
\newblock {\em Information Sciences}, 2013.

\bibitem{ziegler1942optimum}
J.~G. Ziegler and N.~B. Nichols.
\newblock Optimum settings for automatic controllers.
\newblock {\em Trans. ASME}, 1942.

\bibitem{hjalmarsson2002iterative}
H.~Hjalmarsson.
\newblock Iterative feedback tuning—an overview.
\newblock {\em International journal of adaptive control and signal
  processing}, 2002.

\bibitem{karimi2002convergence}
A.~Karimi, L.~Mi{\v{s}}kovi{\'c}, and D.~Bonvin.
\newblock Convergence analysis of an iterative correlation-based controller
  tuning method.
\newblock {\em IFAC Proceedings Volumes}, 2002.

\bibitem{campi2002virtual}
M.~C. Campi, A.~Lecchini, and S.~M. Savaresi.
\newblock Virtual reference feedback tuning: a direct method for the design of
  feedback controllers.
\newblock {\em Automatica}, 2002.

\bibitem{kergus2017frequency}
P.~Kergus, C.~Poussot-Vassal, F.~Demourant, and S.~Formentin.
\newblock Frequency-domain data-driven control design in the loewner framework.
\newblock {\em IFAC-PapersOnLine}, 2017.

\bibitem{bazanella2011data}
AS. Bazanella, L.~Campestrini, and D.~Eckhard.
\newblock {\em Data-driven controller design: the H2 approach}.
\newblock Springer, 2011.

\bibitem{de1999iterative}
F.~De~Bruyne and L.~C. Kammer.
\newblock Iterative feedback tuning with guaranteed stability.
\newblock In {\em American Control Conference}. IEEE, 1999.

\bibitem{karimi2007non}
A.~Karimi, K.~Van~Heusden, and D.~Bonvin.
\newblock Non-iterative data-driven controller tuning using the correlation
  approach.
\newblock In {\em European Control Conference}. IEEE, 2007.

\bibitem{selvi2018towards}
D.~Selvi, D.~Piga, and A.~Bemporad.
\newblock Towards direct data-driven model-free design of optimal controllers.
\newblock In {\em European Control Conference}. IEEE, 2018.

\bibitem{van2009data}
K.~van Heusden, A.~Karimi, and D.~Bonvin.
\newblock Data-driven controller validation.
\newblock In {\em 15th IFAC Symposium on System Identification}, 2009.

\bibitem{havre1997limitations}
K.~Havre and S.~Skogestad.
\newblock Limitations imposed by rhp zeros/poles in multivariable systems.
\newblock In {\em European Control Conference}. IEEE, 1997.

\bibitem{rojas2012analyzing}
C.~R. Rojas, T.~Oomen, H.~Hjalmarsson, and B.~Wahlberg.
\newblock Analyzing iterations in identification with application to
  nonparametric h$\infty$-norm estimation.
\newblock {\em Automatica}, 2012.

\bibitem{rallo2017data}
G.~Rallo, S.~Formentin, C.~R. Rojas, T.~Oomen, and S.~M. Savaresi.
\newblock Data-driven h$\infty$-norm estimation via expert advice.
\newblock In {\em 56th Annual Conference on Decision and Control}. IEEE, 2017.

\bibitem{mayo2007framework}
A.~J. Mayo and A.~C. Antoulas.
\newblock A framework for the solution of the generalized realization problem.
\newblock {\em Linear algebra and its applications}, 2007.

\bibitem{antoulas2005approximation}
A.~C. Antoulas.
\newblock {\em Approximation of large-scale dynamical systems}.
\newblock SIAM, 2005.

\bibitem{stevens2015aircraft}
B.~L. Stevens, F.~L. Lewis, and E.~N. Johnson.
\newblock {\em Aircraft control and simulation: dynamics, controls design, and
  autonomous systems}.
\newblock Wiley, 2015.

\end{thebibliography}
\end{document}